\documentclass[sigconf,natbib=true]{acmart}
\usepackage{amsmath}
\usepackage{booktabs}
\usepackage{threeparttable}
\usepackage{multirow}
\usepackage[inline]{enumitem}
\usepackage{subcaption}
\usepackage{amsthm}
\usepackage{enumitem}
\usepackage{tikz}
\usepackage{amsfonts}
\usepackage{algorithm}
\usepackage[noend]{algpseudocode}
\usetikzlibrary{bayesnet}
\usepackage{xcolor}
\usepackage[utf8]{inputenc}
\usepackage{arydshln}

\newcommand{\bb}[1]{\mathbf{#1}}

\newcommand{\br}{\bb{r}}
\newcommand{\bhr}{\bb{\hat{r}}}
\newcommand{\bR}{\bb{R}}

\newcommand{\bhRzero}{\bb{\hat{R}}^0}
\newcommand{\bhRone}{\bb{\hat{R}}^1}
\newcommand{\bru}{\br_{u}}
\newcommand{\bhru}{\bhr_{u}}
\newcommand{\bhrut}{\bhr_{u}^{t}}

\newcommand{\brui}{\br_{u,i}}

\newcommand{\bkp}{\bb{k}^{+}}
\newcommand{\bkn}{\bb{k}^{-}}
\newcommand{\bhk}{\bb{\hat{k}^+}}
\newcommand{\bhnk}{\bb{\hat{k}^-}}
\newcommand{\bKp}{\bb{K}^{+}}
\newcommand{\bKn}{\bb{K}^{-}}
\newcommand{\bKI}{\bb{K^I}}
\newcommand{\bkpu}{\bkp_{u}}

\newcommand{\bknu}{\bkn_{u}}

\newcommand{\bkIi}{\bb{k}^{I}_{i}}
\newcommand{\bhku}{\bhk_{u}}
\newcommand{\bhnku}{\bhnk_{u}}

\newcommand{\bz}{\bb{z}}

\newcommand{\bzu}{\bz_{u}}

\newcommand{\bzcpt}{\bz_{{c_u^+}^{t}}}
\newcommand{\bzcnt}{\bz_{{c_u^-}^{t}}}

\newcommand{\bc}{\bb{c}^{+}}
\newcommand{\bcn}{\bb{c}^{-}}

\newcommand{\bcut}{{\bc_{u}}^{t}}
\newcommand{\bcnut}{{\bcn_{u}}^{t}}

\newcommand*\circled[1]{\tikz[baseline=(char.base)]{
            \node[shape=circle,draw,inner sep=0.25pt] (char) {{\normalfont #1}};}}
            
\makeatletter
\def\adl@drawiv#1#2#3{%
        \hskip.5\tabcolsep
        \xleaders#3{#2.5\@tempdimb #1{1}#2.5\@tempdimb}%
                #2\z@ plus1fil minus1fil\relax
        \hskip.5\tabcolsep}
\newcommand{\cdashlinelr}[1]{%
  \noalign{\vskip\aboverulesep
           \global\let\@dashdrawstore\adl@draw
           \global\let\adl@draw\adl@drawiv}
  \cdashline{#1}
  \noalign{\global\let\adl@draw\@dashdrawstore
           \vskip\belowrulesep}}
\makeatother

\acmConference[arXiv]{arXiv}{2022}{USA}
\acmBooktitle{arXiv}
\setcopyright{acmcopyright}
\copyrightyear{2022}
\acmYear{2022}
\acmDOI{ }
\begin{document}

\title{Positive \& Negative Critiquing for VAE-based Recommenders}

\author{Diego Antognini}
\affiliation{
  \institution{
École Polytechnique Fédérale de Lausanne}
  \city{Lausanne}
  \country{Switzerland}
}
\email{diego.antognini@epfl.ch}

\author{Boi Faltings}
\affiliation{
  \institution{
École Polytechnique Fédérale de Lausanne}
  \city{Lausanne}
  \country{Switzerland}
}
\email{boi.faltings@epfl.ch}

\begin{abstract}

Providing explanations for recommended items allows users to refine the recommendations by \textit{critiquing} parts of the explanations.~As a result of revisiting critiquing from the perspective of multimodal generative models, recent work has proposed M\&Ms-VAE, which achieves state-of-the-art performance in terms of recommendation, explanation, and critiquing. M\&Ms-VAE and similar models~allow users to negatively critique (i.e., explicitly disagree). However, they share a significant  drawback: users cannot positively critique (i.e., highlight a desired feature).
We address this deficiency with M\&Ms-VAE\textsuperscript{$+$}, an extension of M\&Ms-VAE that enables positive~and negative critiquing. In addition to modeling users' interactions and keyphrase-usage preferences, we model their keyphrase-usage dislikes. Moreover, we design a novel critiquing module that is trained in a self-supervised fashion.
Our experiments on two datasets show that M\&Ms-VAE\textsuperscript{$+$} matches or exceeds M\&Ms-VAE in recommendation and explanation performance. Furthermore, our results demonstrate that representing positive and negative critiques differently enables M\&Ms-VAE\textsuperscript{$+$} to significantly outperform M\&Ms-VAE and other models in positive \textit{and} negative multi-step critiquing.

\end{abstract}

\maketitle

\section{Introduction}

Critiquing is a conversational recommendation method that incrementally adapts recommendations in response to user preferences \cite{chen2012critiquing}. Several studies have revisited critiquing with neural models~\cite{keyphraseExtractionDeep,antognini2020interacting,luo2020,luo2020b,hanze2020}. They allow users to critique the recommendation by~interacting with a set of attributes mined from user reviews.
Recently, \cite{fast_critiquing} proposed a multimodal variational-autoencoder, called M\&Ms-VAE, that achieves state-of-the-art performance in terms of recommendation, explanation, and multi-step critiquing. Although M\&Ms-VAE has made important contributions, it shares an important deficiency with other models: users can only express an explicit disagreement (i.e., negative critiquing); there is no means of highlighting a desired feature (i.e., positive critiquing) \cite{multi_step_demo,sanner:sigir21}. 

One way to enable positive critiquing is to embed the positive critique identically to the negative one, with the advantage that~the recommender remains the same. As we demonstrate empirically in Section~\ref{sec_rq2}, it underperforms in positive multi-step critiquing compared~to negative critiquing.
Another means of enabling positive critiquing consists of modeling the positive and negative critiques within the recommender. Although this approach improves critiquing performance, it affects recommendation and explanation performance,  in a way that limits their practice performance.

In this study, we present M\&Ms-VAE\textsuperscript{$+$}, an extension of M\&Ms-VAE, to enable positive critiquing. We take advantage of the multimodal modeling to introduce a third modality that represents~users' keyphrase-usage dislikes along with their interactions and keyphrase-usage preferences. Finally, we design a method to critique positively and negatively, which trained in a self-supervision manner.

We evaluate our method using two real-world, publicly available datasets. The results show that M\&Ms-VAE\textsuperscript{$+$} \begin{enumerate*}
 \item matches or exceeds M\&Ms-VAE in recommendation and explanation performance and
 \item outperforms M\&Ms-VAE and other strong models by a large margin on positive and negative multi-step critiquing.
 \end{enumerate*}

\section{MIXTURE-OF-EXPERTS MULTIMODAL~VAE}
\label{sec_mms_vae}
\paragraph{Notation.}Before proceeding, we define the following notation:% used throughout this paper:
\begin{itemize}
	\item $U$, $I$, and $K$: The user, item, and keyphrase sets, respectively.
	\item $\bR \in \mathbb{R}^{|U|\times|I|}$: The binary user-by-item interaction matrix.
	\item $\bKp, \bKn \in \mathbb{R}^{|U|\times|K|}$: User-keyphrase matrices that reflect users' keyphrase-usage preferences and dislikes in the reviews. $\bkpu$ are the keywords mentioned by the user $u$, and~$\bknu$ are all the others. Similarly, $\bKI$ is the item-keyphrase matrix.
	%\item $\bKI \in \mathbb{R}^{|I|\times|K|}$: The binary item-keyphrase matrix.
	%\item $\bzu \in \mathbb{R}^{|H|}$: The user $u$'s latent embedding of dimension $H$.
	\item $\bcut, \bcnut \in \mathbb{R}^{|K|}$: A one-hot vector whose only positive value indicates the index of the keyphrase to be critiqued positively and negatively by the user $u$ at a given step $t$.
	\item $\bzu, \bzcpt, \bzcnt \in \mathbb{R}^{|H|}$: The latent representation of the user~$u$, the positive critique $\bcut$, and the negative critique $\bcnut$.
\end{itemize}
\paragraph{Model Overview.}An important feature of M\&Ms-VAE~\cite{fast_critiquing} is learning the joint distribution $p(\bru, \bkpu)$ under partial observations: we aim to recommend and generate keyphrase explanation jointly and independently from each observed variable (i.e., modality). Thanks to the multimodal assumption, $\bru$~and $\bkpu$ are~conditionally independent given the common latent variable $\bzu$; an unobserved variable can be safely ignored when evaluating the marginal likelihood As Fig.~\ref{fig_gm} (left) shows, we write the joint log-likelihood as follows:\begin{align}
	\label{eq_mm}
	\log~p(\bru, \bkpu) &\ge \mathbb{E}_{q_\Phi(\bzu | \bru, \bkpu)} \bigl[\log p_{\Theta_r}(\bru | \bzu) + \log p_{\Theta_{k^{+}}}(\bkpu | \bzu)\bigr] \nonumber\\  &- \beta \textrm{ D}_{\textrm{KL}} \bigl[ q_\Phi(\bzu | \bru, \bkpu) ~||~ p(\bzu) \bigr],
\end{align}where the prior $p(\bzu)$ is a normal distribution and $\beta$ is a hyperparameter that controls the strength of the regularization relative.

Currently, learning $q_\Phi(\bzu | \bru, \bkpu)$ requires $\bru$ and $\bkpu$ to be fully observed. \cite{fast_critiquing} remedied this problem by using a mixture of experts (MoE): $q_\Phi(\bzu | \bru, \bkpu) = \zeta\bigl(q_{\Phi_r}(\bzu | \bru), q_{\Phi_{k^+}}(\bzu | \bkpu)\bigr)$. The MoE $\zeta(\cdot)$ takes the average of the experts when $\bru$ and $\bkpu$ are both observed or the unimodal expert when only one modality is observed. 

Finally, the proposed training scheme mimics weakly supervised learning to train the individual inference networks $q_{\Phi_r}$ and $q_{\Phi_{k^+}}$:\begin{equation}
	\label{eq_mm_final}
	\mathcal{L}_{M\&Ms}(\bR, \bKp) = \sum\nolimits_{u} ELBO(\bru, \bkpu) + ELBO(\bru) + ELBO(\bkpu)
\end{equation}

\subsection{Negative Keyphrase-based Critiquing}
\label{sec_neg_critiquing}
\begin{algorithm}[!t]
\caption{\label{alg_dataset}Creation of Synthetic Critiquing Datasets}
\begin{algorithmic}[1]
\Function{Create\_Critiquing\_Dataset}{$K$, $\bR^{\textrm{val}}$, $\bKI$, $\bb{\hat{K}^U}$}
	\State Synthetic datasets $D^{+} \leftarrow \{\}, D^{-} \leftarrow \{\}$
    \For{each user $u$ and target item $i$, where $\brui^{\textrm{val}}=1$}
        \State Randomly sample a critique $\bcn \in K \backslash \bkIi$ \textcolor{blue}{\textit{// Neg. critique}}
        \State Compute the (un-)affected item sets $I^{-}_{\bcn}$ and $I^{+}_{\bcn}$ by $\bcn$ 
        \State Update $D^{-} \leftarrow D^{-} \cup \{(u, i, \bcn, I^{+}_{\bcn}, I^{-}_{\bcn})\}$
        \State Randomly sample a critique $\bc \in \bkIi \backslash \bhku$ \textcolor{blue}{\textit{// Pos. critique}}
        \State Compute the (un-)affected item sets $I^{+}_{\bc}$ and $I^{-}_{\bc}$ by $\bc$ 
        \State Update $D^{+} \leftarrow D^{+} \cup \{(u, i, \bc, I^{+}_{\bc}, I^{-}_{\bc})\}$
        \EndFor
    \State \textbf{return} Synthetic datasets $D^{-}$ and $D^{+}$
\EndFunction
\end{algorithmic}
\end{algorithm}
Given the predicted explanation $\bhku$ and the recommendation $\bhru$, the user $u$ can accept or refine the recommendation. In negative~critiquing, the user iteratively provides a keyphrase to critique~$\bcnut$ (i.e., disagreement) and obtains a new recommendation $\bhrut$ until he is satisfied. 
The critique representation $\bzcnt$ is encoded via $q_{\Phi_{k^+}}(\bzu | \bkpu)$.

To blend the user representation $\bzu$ with the $t$\textsuperscript{th} critique representation $\bzcnt$, \cite{fast_critiquing} introduced a blending function that treats each critique as independent and uses gated recurrent units~\cite{chung2014empirical}:~$\xi(\bzu, \bzcnt)$ $= \textrm{GRU}([\bzu; \bzcnt])$, where \textit{;} denotes the concatenation operation.
The blending module $\xi(\cdot)$ is optimized in a self-supervised fashion. First, a synthetic dataset $D^-$ is created following~Alg.~\ref{alg_dataset} (Lines 1-6 and Line 10). For each negative critique $\bcn$ inconsistent with the target item, we compute the item set $I^{+}_{\bcn}$ (symmetrically $I^{-}_{\bcn}$), which contains (symmetrically does not contain) the~critique. Second, a max-margin ranking-based objective encourages $\xi(\cdot)$ to learn how to re-rank the items according to the critique $\bcn$: \begin{align}
	\label{eq_ssc}
	&\mathcal{L}_{M\&Ms}^{neg}(\bhRzero,\bhRone,u,\bcn,I^{+}_{\bcn},I^{-}_{\bcn}) = \sum\nolimits_{i^+ \in I^{+}_{\bcn}} \max \{0, h - (\bhr_{u,i^+}^{0} - \bhr_{u,i^+}^{1}) \}\nonumber\\ & + \sum\nolimits_{i^- \in I^{-}_{\bcn}} \max \{0, h - (\bhr_{u,i^-}^{1} - \bhr_{u,i^-}^{0}) \}.
\end{align}
\section{Enabling Positive critiquing}
\label{sec_positive_critiquing}

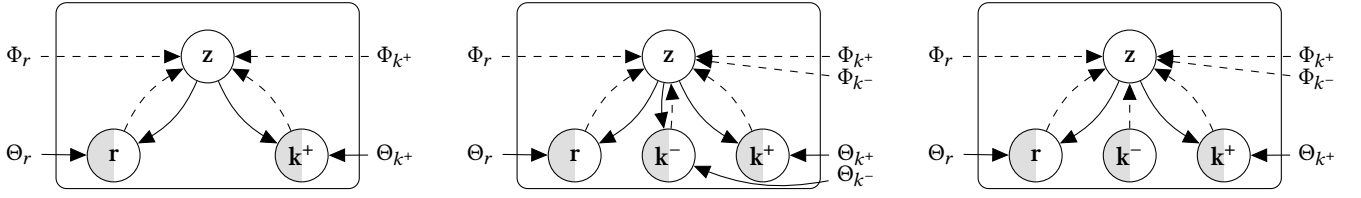
\begin{figure*}[]
\centering
\begin{minipage}[t]{0.3\textwidth}
  \centering
  \begin{tikzpicture}
  	  \node[latent] (z) {$\bz$};
      \node[latent, below=of z, xshift=-1.25cm,yshift=0.4cm,path picture={\fill[gray!25] (path picture bounding box.south) rectangle (path picture bounding box.north west);}] (r) {$\br$} ;
      \node[latent, below=of z, xshift=1.25cm,yshift=0.4cm,path picture={\fill[gray!25] (path picture bounding box.south) rectangle (path picture bounding box.north west);}] (k) {$\bkp$} ;      

	  \node[xshift=2.5cm] (1) {$\Phi_{k^+}$};
	  \node[xshift=-2.5cm] (2) {$\Phi_r$};
	  \node[below=of z, yshift=0.3cm, xshift=2.5cm] (3) {$\Theta_{k^+}$};
	  \node[below=of z, yshift=0.3cm, xshift=-2.5cm] (4) {$\Theta_r$};
    \plate[inner xsep=0.4cm, inner ysep=0.1cm, xshift=0cm, yshift=0.25cm] {plate} {(r) (k) (z)} {};
    
    \node[fill, color=white, opacity=0] at (-1.03cm,0.54cm) {\textbf{M\&Ms-VAE}};
    \node[fill, color=white, opacity=0] at (1.05cm,0.5cm) {$u \in \{1 \dots U\}$};
    
	\path[->] (z) edge [bend left=20] node {} (r) ;
	\path[->] (z) edge [bend left=-20] node {} (k) ;
	\path[->, dashed] (r) edge [bend left=20] node {} (z) ;
	\path[->, dashed] (k) edge [bend left=-20] node {} (z) ;
	
	\path[->, dashed] (1) edge [] node {} (z) ;
	\path[->, dashed] (2) edge [] node {} (z) ;
	\path[->] (3) edge [] node {} (k) ;
	\path[->] (4) edge [] node {} (r) ;
  \end{tikzpicture}
\end{minipage}
\hspace{2em}
\begin{minipage}[t]{0.3\textwidth}
  \centering
  \begin{tikzpicture}
  	  \node[latent] (z) {$\bz$};
      \node[latent, below=of z, xshift=-1.25cm,yshift=0.4cm,path picture={\fill[gray!25] (path picture bounding box.south) rectangle (path picture bounding box.north west);}] (r) {$\br$} ;
      \node[latent, below=of z, xshift=1.25cm,yshift=0.4cm,path picture={\fill[gray!25] (path picture bounding box.south) rectangle (path picture bounding box.north west);}] (k) {$\bkp$} ;      
      \node[latent, below=of z,yshift=0.4cm,path picture={\fill[gray!25] (path picture bounding box.south) rectangle (path picture bounding box.north west);}] (kn) {$\bkn$} ;
      
	  \node[xshift=2.5cm] (1) {$\Phi_{k^+}$};
	  \node[xshift=2.5cm, yshift=-0.3cm] (5) {$\Phi_{k^-}$};
	  \node[xshift=-2.5cm] (2) {$\Phi_r$};
	  \node[below=of z, yshift=0.3cm, xshift=2.5cm] (3) {$\Theta_{k^+}$};
	  \node[below=of z, yshift=0.3cm, xshift=-2.5cm] (4) {$\Theta_r$};
	  \node[below=of z, yshift=0.025cm, xshift=2.5cm] (6) {$\Theta_{k^-}$};
    \plate[inner xsep=0.4cm, inner ysep=0.1cm, xshift=0cm, yshift=0.25cm] {plate} {(r) (k) (z)} {};
    \node[fill, color=white, opacity=0] at (-0.95cm,0.5cm) {\textbf{M\&Ms-VAE\textsuperscript{3}}};
    \node[fill, color=white, opacity=0] at (1.05cm,0.5cm) {$u \in \{1 \dots U\}$};
    
	\path[->] (z) edge [bend left=20] node {} (r) ;
	\path[->] (z) edge [bend left=-20] node {} (k) ;
	\path[->] (z) edge [bend left=-10] node {} (kn) ;
	\path[->, dashed] (r) edge [bend left=20] node {} (z) ;
	\path[->, dashed] (k) edge [bend left=-20] node {} (z) ;
	\path[->, dashed] (kn) edge [bend left=-5] node {} (z) ;
	\path[->] (6) edge [bend left=16.5] node {} (kn) ;
	
	\path[->, dashed] (1) edge [] node {} (z) ;
	\path[->, dashed] (2) edge [] node {} (z) ;
	\path[->, dashed] (5) edge [] node {} (z) ;
	\path[->] (3) edge [] node {} (k) ;
	\path[->] (4) edge [] node {} (r) ;
  \end{tikzpicture}
\end{minipage}
\hspace{2em}
\begin{minipage}[t]{0.3\textwidth}
  \centering
  \begin{tikzpicture}
  	  \node[latent] (z) {$\bz$};
      \node[latent, below=of z, xshift=-1.25cm,yshift=0.4cm,path picture={\fill[gray!25] (path picture bounding box.south) rectangle (path picture bounding box.north west);}] (r) {$\br$} ;
      \node[latent, below=of z, xshift=1.25cm,yshift=0.4cm,path picture={\fill[gray!25] (path picture bounding box.south) rectangle (path picture bounding box.north west);}] (k) {$\bkp$} ;      
      \node[latent, below=of z, yshift=0.4cm,path picture={\fill[gray!25] (path picture bounding box.south) rectangle (path picture bounding box.north west);}] (kn) {$\bkn$} ;
      
	  \node[xshift=2.5cm] (1) {$\Phi_{k^+}$};
	  \node[xshift=2.5cm, yshift=-0.3cm] (5) {$\Phi_{k^-}$};
	  \node[xshift=-2.5cm] (2) {$\Phi_r$};
	  \node[below=of z, yshift=0.3cm, xshift=2.5cm] (3) {$\Theta_{k^+}$};
	  \node[below=of z, yshift=0.3cm, xshift=-2.5cm] (4) {$\Theta_r$};
    \plate[inner xsep=0.4cm, inner ysep=0.1cm, xshift=0cm, yshift=0.25cm] {plate} {(r) (k) (z)} {};
    \node[fill, color=white, opacity=0] at (-0.95cm,0.5cm) {\textbf{M\&Ms-VAE\textsuperscript{$+$}}};
    \node[fill, color=white, opacity=0] at (1.05cm,0.5cm) {$u \in \{1 \dots U\}$};
    
	\path[->] (z) edge [bend left=20] node {} (r) ;
	\path[->] (z) edge [bend left=-20] node {} (k) ;
	\path[->, dashed] (r) edge [bend left=20] node {} (z) ;
	\path[->, dashed] (k) edge [bend left=-20] node {} (z) ;
	\path[->, dashed] (kn) edge [bend left=0] node {} (z) ;
	
	\path[->, dashed] (1) edge [] node {} (z) ;
	\path[->, dashed] (2) edge [] node {} (z) ;
	\path[->, dashed] (5) edge [] node {} (z) ;
	\path[->] (3) edge [] node {} (k) ;
	\path[->] (4) edge [] node {} (r) ;
  \end{tikzpicture}
\end{minipage}
    \caption{\label{fig_gm}Probabilistic-graphical-model views of M\&Ms-VAE \cite{fast_critiquing} (left), M\&Ms-VAE\textsuperscript{3} (middle), and M\&Ms-VAE\textsuperscript{$+$} (right). In all models, both the implicit feedback $\bru$ and the keyphrase-usage preferences $\bkpu$ are generated from user $u$'s latent representation $\bzu$. Regarding the keyphrase-usage~dislikes $\bknu$, M\&Ms-VAE\textsuperscript{3} considers $\bknu$ a third modality whereas M\&Ms-VAE\textsuperscript{$+$} treats it only as an input variable. Solid lines denote the generative model, whereas dashed lines denote the variational approximation.}
\end{figure*}

A drawback of M\&Ms-VAE and other models (\cite{keyphraseExtractionDeep,luo2020,luo2020b,hanze2020})~is that users can express a disagreement but cannot highlight a feature. To enable positive critiquing, we propose three incremental solutions.

\paragraph{\protect\circled{1} Extending the blending module.} A suboptimal way to enable positive critiquing is to adjust \textit{only} the blending module and~the self-supervised task. The advantage of this approach is that there is no need to retrain the generative and inference networks: the recommendation and explanation performance remain the same. The representation $\bzcpt$ of the positive critique $\bcut$ is computed~identically to $\bzcnt$, using the inference model $q_{\Phi_{k^+}}(\bzu | \bkpu)$.
Then, we decompose the blending module into $\xi^+(\bzu,\bzcpt) = \textrm{GRU}([\bb{e^+}; \bzu;$ $\bzcpt])$ and $\xi^-(\bzu,\bzcnt)$ $= \textrm{GRU}([\bb{e^-}; \bzu; \bzcnt])$. $\xi^+(\cdot)$ and $\xi^-(\cdot)$ share the same weights. Both $\bb{e^+}$ and $\bb{e^-}$ are trainable parameters that condition the gated mechanism to compute a representation that reflects an explicit agreement with $\bcut$ and a disagreement~with~$\bcnut$.

To include positive critiquing examples, we create a second synthetic dataset $D^+$ (see Alg.~\ref{alg_dataset}). Echoing the approach used in Section~\ref{sec_neg_critiquing}, we sample a critique $\bc$ that is not part of the predicted keyphrase explanation $\bhku$ but is consistent with the target item's~features and compute the item sets $I^{+}_{\bc}$ and $I^{-}_{\bc}$. \begin{align}
	\label{eq_ssc_pos_neg}
	&\mathcal{L}_{M\&Ms}^{pos,neg}(\cdot) = \mathcal{L}_{M\&Ms}^{neg}(\cdot) + \sum\nolimits_{i^+ \in I^{+}_{\bc}} \max \{0, h - (\bhr_{u,i^+}^{1} - \bhr_{u,i^+}^{0}) \} \nonumber \\
	&+\sum\nolimits_{i^- \in I^{-}_{\bc}} \max \{0, h - (\bhr_{u,i^-}^{0} - \bhr_{u,i^-}^{1}) \}.
\end{align}
\paragraph{\protect\circled{2} Introducing a third modality.} Above, we used the \textit{same} model $q_{\Phi_{k^+}}(\cdot)$ to infer $\bzcpt$ and $\bzcnt$. However, neither $\bc$ and $\bcn$ embed the same meaning, whereas $q_{\Phi_{k^+}}(\cdot)$ considers at training only users' keyphrase-usage preferences. Therefore, we introduce a new modality $\bknu$ that represents users' keyphrase-usage~dislikes: the keyphrases that are not part of the user's profile.
As Fig.~\ref{fig_gm} (M\&Ms-VAE\textsuperscript{3}) shows, we update Eq.~\ref{eq_mm} as follows: \begin{align}
\label{eq_mm3_ll}
&\log~p(\bru, \bkpu, \bknu) \ge \mathbb{E}_{q_\Phi(\bzu | \bru, \bkpu, \bknu)} \bigl[\log p_{\Theta_r}(\bru | \bzu) + \log p_{\Theta_{k^{+}}}(\bkpu | \bzu) \nonumber \\& + \log p_{\Theta_{k^{-}}}(\bknu | \bzu)\bigr] - \beta \textrm{ D}_{\textrm{KL}} \bigl[ q_\Phi(\bzu | \bru, \bkpu, \bknu) ~||~ p(\bzu) \bigr].
\end{align}
We augment the mixture of experts with the extra unimodal posterior: $q_\Phi(\bzu | \bru, \bkpu, \bknu) = \zeta\bigl(q_{\Phi_r}(\bzu | \bru), q_{\Phi_{k^+}}(\bzu | \bkpu), q_{\Phi_{k^-}}(\bzu | \bknu)\bigr)$. We also update the training strategy as $\mathcal{L}_{M\&Ms^3}(\bR, \bKp, \bKn)$: \begin{equation}
	\label{eq_mm3_final}
	 \sum\nolimits_{u} ELBO(\bru, \bkpu, \bknu) + ELBO(\bru) + ELBO(\bkpu) + ELBO(\bknu).
\end{equation}
Finally, we compute the critique representations $\bzcpt$ and $\bzcnt$ using $q_{\Phi_{k^+}}(\bzu | \bkpu)$ and $q_{\Phi_{k^-}}(\bzu | \bknu)$. 
However, the joint likelihood now includes the generative model $p_{\Theta_{k^-}}(\bknu | \bzu)$, which leads to~two problems: \begin{enumerate*}
 \item it is unclear what value $\bhnku$ offers the user in addition to the explanation $\bhku$ and the recommendation $\bhru$, and 
 \item $\bhnku$ is redundant, because it is equivalent to $K \backslash \bhku$. It might affect~the~latent space and the recommendation and explanation performance.
 \end{enumerate*}
\paragraph{\protect\circled{3} Removing $p_{\Theta_{k^-}}(\bknu | \bzu)$.}
Because we aim to develop a generative model of the form $p_\Theta(\bru, \bkpu, \bzu)$, and thanks to the multimodal factorization, we can rewrite Eq.~\ref{eq_mm3_ll} by safely removing $p_{\Theta_{k^-}}(\bknu | \bzu)$: \begin{align}
\label{eq_mm_3}
\log~p(\bru, \bkpu) &\ge \mathbb{E}_{q_\Phi(\bzu | \bru, \bkpu, \bknu)} \bigl[\log p_\Theta(\bru,| \bzu) + \log p_{\Theta_{k^{+}}}(\bkpu | \bzu)\bigr] \nonumber \\ &- \beta \textrm{ D}_{\textrm{KL}} \bigl[ q_\Phi(\bzu | \bru, \bkpu, \bknu) ~||~ p(\bzu) \bigr].
\end{align}To summarize, we model both keyphrase-usage preferences and dislikes of users, while predicting only the keyphrase explanation alongside the recommended items, as Fig.~\ref{fig_gm} (M\&Ms-VAE\textsuperscript{$+$}) shows. The inference networks are identical to those of M\&Ms-VAE\textsuperscript{3} and similar to the generative networks of M\&Ms-VAE. A crucial question remains: how do we train $q_{\Phi_{k^-}}(\bzu | \bknu)$? Once again, we take advantage of the multimodal modeling and incorporate the partial-paired observations $(\bru,\bknu)$ and $(\bkpu,\bknu)$ into the training scheme: \begin{align}
	\label{eq_mmp_final}
	&\mathcal{L}_{M\&Ms^{+}}(\bR, \bKp, \bKn) = \sum\nolimits_{u \in U} ELBO(\bru, \bkpu, \bknu) \\
	&  + ELBO(\bru)+ ELBO(\bkp) + ELBO(\bru, \bknu) + ELBO(\bkp, \bknu).\nonumber 
\end{align}

\section{Experiments}

\subsection{Datasets}
\label{sec_datasets}
\begin{table}[!t]
%\small
    \centering
   \caption{\label{stats_datasets}Descriptive statistics of the datasets.}
\begin{threeparttable}
\begin{tabular}{@{}l@{\hspace{1mm}}c@{\hspace{1mm}}c@{\hspace{1mm}}c@{\hspace{1mm}}c@{\hspace{1mm}}c@{}}
\textbf{Dataset} & \textbf{\#Users} & \textbf{\#Items} & \textbf{\#Interactions} & \textbf{Sparsity} & \textbf{\#Keyphrases}\\
\toprule
Yelp & 9,801 & 4,706 & 140,496 & 99.70\% & 234\\
Hotel & 7,044 & 4,874 & 143,612 & 99.58\% & 141
\end{tabular}
\end{threeparttable}
\end{table}
We run experiments on two real-world datasets: Yelp~\cite{yelpdataset} and HotelRec~\cite{antognini-faltings-2020-hotelrec}. Each contains over 100k reviews with five-star ratings. As in \cite{fast_critiquing,luo2020,hanze2020}, we extract the positive keyphrases from user reviews for the explanations. We consider keyphrases negative when they are not mentioned by users in their reviews. Each dataset contains complete observations and is split into~60\% 20\%/20\% for the train, dev, and test sets. We binarize the ratings with the thresholds $t > 4.5$ and $t > 3.5$ for restaurants and hotels, respectively.
\subsection{Experimental Settings}

We treat the prior and the likelihood as normal and multinomial distributions. Each inference and generative network is composed of a two-layer neural network with a tanh activation function. We use dropout~\cite{srivastava2014dropout} and the Adam optimizer~\cite{KingmaB14} with a learning rate of $5\cdot10^{-5}$. We anneal linearly the regularization parameter $\beta$ of the KL terms. We tune each model on the NDCG on the validation set with a random search and a maximum of 50 trials. For critiquing, we tune the M\&Ms-VAE-based models on the synthetic datasets.

\subsection{RQ 1: Recommendation and Explanation Performance Comparison}

\subsubsection{Baselines}
\label{sec_rq1_baselines}
We compare M\&Ms-VAE\textsuperscript{$+$} with the state-of-the-art M\&Ms-VAE model, M\&Ms-VAE\textsuperscript{3} (see Section~\ref{sec_positive_critiquing}), and the following baselines.\textbf{PLRec}~\cite{sedhain2016practical} is a linear recommender that solves the scalability problem; it projects the user preferences into smaller space prior to a linear regression. \textbf{PureSVD}~\cite{10.1145/1864708.1864721} builds a similarity matrix through SVD decomposition of the implicit rating matrix. \textbf{CE-VAE}~\cite{luo2020} is a major improvement of the neural collaborative filtering model~\cite{he2017neural} with an explanation and a critiquing module\footnote{We remove the prior model CE-VNCF~\cite{keyphraseExtractionDeep} because it consistently underperformed.}.

\subsubsection{Top-N Recommendation Performance}
\label{sec_rq1_rec} We report the following five metrics: R-Precision and NDCG; MAP,~Precision, and Recall at 10. We present the main results in Table~\ref{table_rec_exp_perf} (left columns).

Interestingly, M\&Ms-VAE\textsuperscript{3} and M\&Ms-VAE\textsuperscript{$+$} perform similarly to M\&Ms-VAE, and the three models obtain the best results. This validates that their modifications do not affect recommendation~performance (M\&Ms-VAE\textsuperscript{$+$} even performs slightly better, particularly on the Hotel dataset).
On both datasets, M\&Ms-VAE-based models significantly outperform CE-VAE by an average factor of~1.7.~These results indicate that the multimodal modeling of M\&Ms-VAE-based models and their training strategies are more robust than those of CE-VAE, which learns a bidirectional mapping between the latent space and the keyphrases, which perturbs the training process.
\subsubsection{Top-K Keyphrase-explanation Performance}
We evaluate the models in terms of keyphrase-explanation performance. PLRec and PureSVD do not contain an explanation module. In Table~\ref{table_rec_exp_perf} (right columns), we report NDCG, MAP, Precision, and Recall at 10.

Overall, M\&Ms-VAE and M\&Ms-VAE\textsuperscript{$+$} obtain the best results. Nevertheless, M\&Ms-VAE\textsuperscript{3} underperforms compared to (respectively outperforms) CE-VAE on the Hotel (respectively Yelp) dataset. It highlights the trade-off between recommendation and explanation that results from the additional training objective or modality.

\begin{table*}[t]
\small
    \centering
\caption{\label{table_rec_exp_perf}Recommendation and keyphrase-explanation results. \textbf{Bold} and \underline{underline} denote the best and second-best results.}
\begin{threeparttable}
\begin{tabular}{@{}cl@{\hspace{1mm}}
c
cc@{\hspace{1mm}}
cc@{\hspace{1mm}}
cc@{\hspace{1mm}}
cc@{\hspace{3mm}}
c@{\hspace{1mm}}
cc@{\hspace{1mm}}
cc@{\hspace{1mm}}
cc@{\hspace{1mm}}
c@{}}
& & \multicolumn{8}{c}{Top-N Recommendation} & & \multicolumn{8}{c}{Top-K Keyphrase Explanation}\\
\cmidrule(lr{1em}){3-10}\cmidrule(lr{1em}){12-19}
& \textbf{Model} & \textbf{R-Precision} & \textbf{NDCG} & & \textbf{MAP@10} & & \textbf{Prec.@10} & & \textbf{Rec.@10} & & \textbf{NDCG@10} & & \textbf{MAP@10} & & \textbf{Prec.@10} & & \textbf{Rec.@10}\\
\toprule
\multirow{6}{*}{\rotatebox{90}{\textit{Yelp}}}
& PLRec & $0.0191$ & $0.0703$ &  & $0.0189$ &  & $0.0166$ &  & $0.0513$ & & - & & - & & - & & -\\
& PureSVD & $0.0253$ & $0.0825$ &  & $0.0249$ &  & $0.0206$ &  & $0.0597$ & & - & & - & & - & & -\\
& CE-VAE & $0.0136$ & $0.0533$ &  & $0.0132$ &  & $0.0119$ &  & $0.0367$ & & $0.2763$ &  & $0.6356$ &  & $0.5876$ &  & $0.1819$\\
& M\&Ms-VAE & $\underline{0.0264}$ & $\mathbf{0.0909}$ &  & $\mathbf{0.0261}$ &  & $\mathbf{0.0223}$ &  & $\underline{0.0682}$ & & $\underline{0.2787}$ &  & $\mathbf{0.6428}$ &  & $\mathbf{0.5935}$ &  & $0.1834$\\
& M\&Ms-VAE\textsuperscript{3} & $0.0263$ & $0.0904$ & & $\underline{0.0259}$ & & $\underline{0.0220}$ & & $\mathbf{0.0685}$ & & $0.2786$ & & $0.6405$ & & $0.5919$ & & $\underline{0.1839}$\\
& M\&Ms-VAE\textsuperscript{$+$} (Ours) & $\mathbf{0.0270}$ & $\underline{0.0905}$ &  & $0.0256$ &  & $0.0216$ &  & $0.0671$ & & $\mathbf{0.2797}$ & & $\underline{0.6413}$ & & $\underline{0.5931}$ & & $\mathbf{0.1842}$\\
\toprule
\multirow{6}{*}{\rotatebox{90}{\textit{Hotel}}}
& PLRec & $0.0242$ & $0.0664$ &  & $0.0234$ &  & $0.0190$ &  & $0.0466$ & & - & & - & & - & & -\\
& PureSVD & $0.0179$ & $0.0541$ &  & $0.0173$ &  & $0.0145$ &  & $0.0357$ & & - & & - & & - & & -\\
& CE-VAE & $0.0147$ & $0.0538$ &  & $0.0146$ &  & $0.0137$ &  & $0.0334$ & & $0.3521$ &  & $0.9113$ &  & $0.8629$ & & $0.2105$\\
& M\&Ms-VAE & $\underline{0.0272}$ & $\mathbf{0.0804}$ &  & $\underline{0.0265}$ &  & $\underline{0.0227}$ &  & $\underline{0.0555}$ & & $\mathbf{0.3595}$ &  & $\mathbf{0.9393}$ &  & $\mathbf{0.8776}$ &  & $\underline{0.2131}$\\
& M\&Ms-VAE\textsuperscript{3} & $0.0262$ & $0.0779$ & & $0.0261$ & & $0.0217$ & & $0.0529$ & & $0.3396$ &  & $0.8668$ &  & $0.8506$ &  & $0.2075$\\
& M\&Ms-VAE\textsuperscript{$+$} (Ours) & $\mathbf{0.0279}$ & $\underline{0.0797}$ &  &  $\mathbf{0.0266}$ &  & $\mathbf{0.0228}$ &  & $\mathbf{0.0561}$ & & $\underline{0.3591}$ & & $\underline{0.9361}$ & & $\mathbf{0.8776}$ & & $\mathbf{0.2133}$\\
\end{tabular}
\end{threeparttable}
\end{table*}
\begin{figure*}[!t]
\centering
\includegraphics[width=0.9\textwidth]{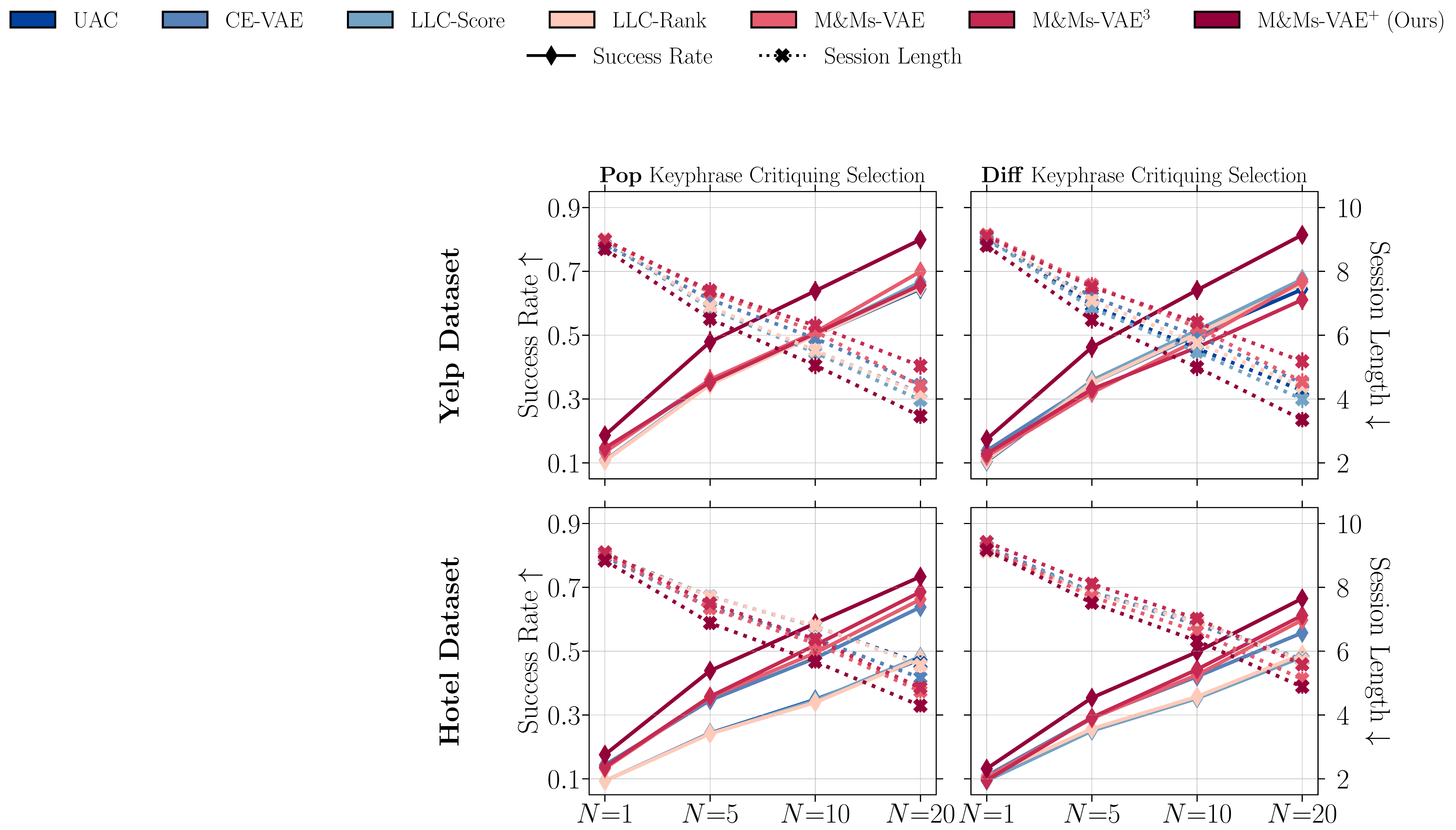}
\\
\begin{subfigure}[t]{0.49\textwidth}
    \centering
    \includegraphics[width=\textwidth,height=5.83336cm]{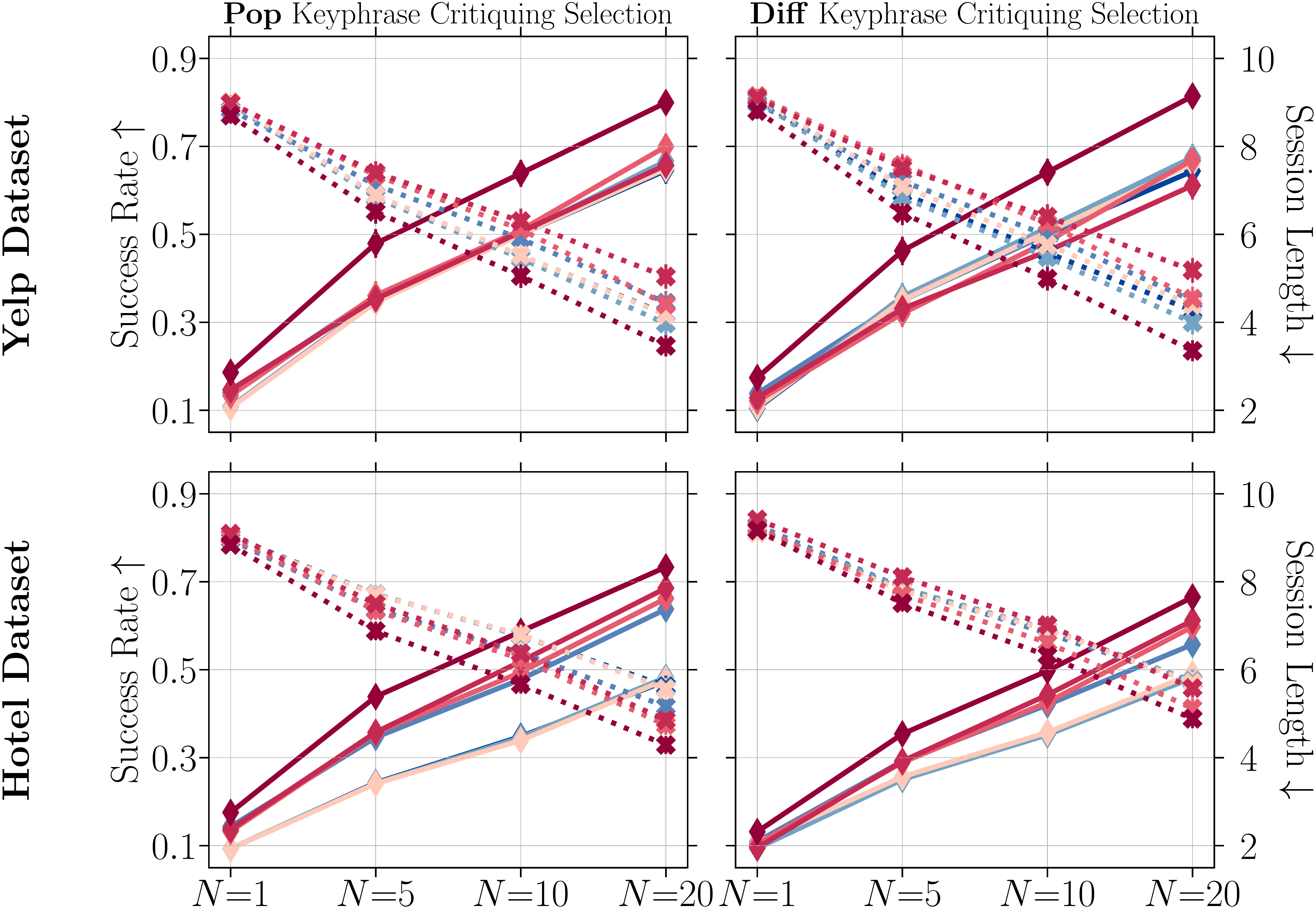}
    \caption{\label{fig_pos_critiquing}Positive critiquing.}
\end{subfigure}
\hfill
\begin{subfigure}[t]{0.49\textwidth}
    \centering
    \includegraphics[width=\textwidth,height=5.83336cm]{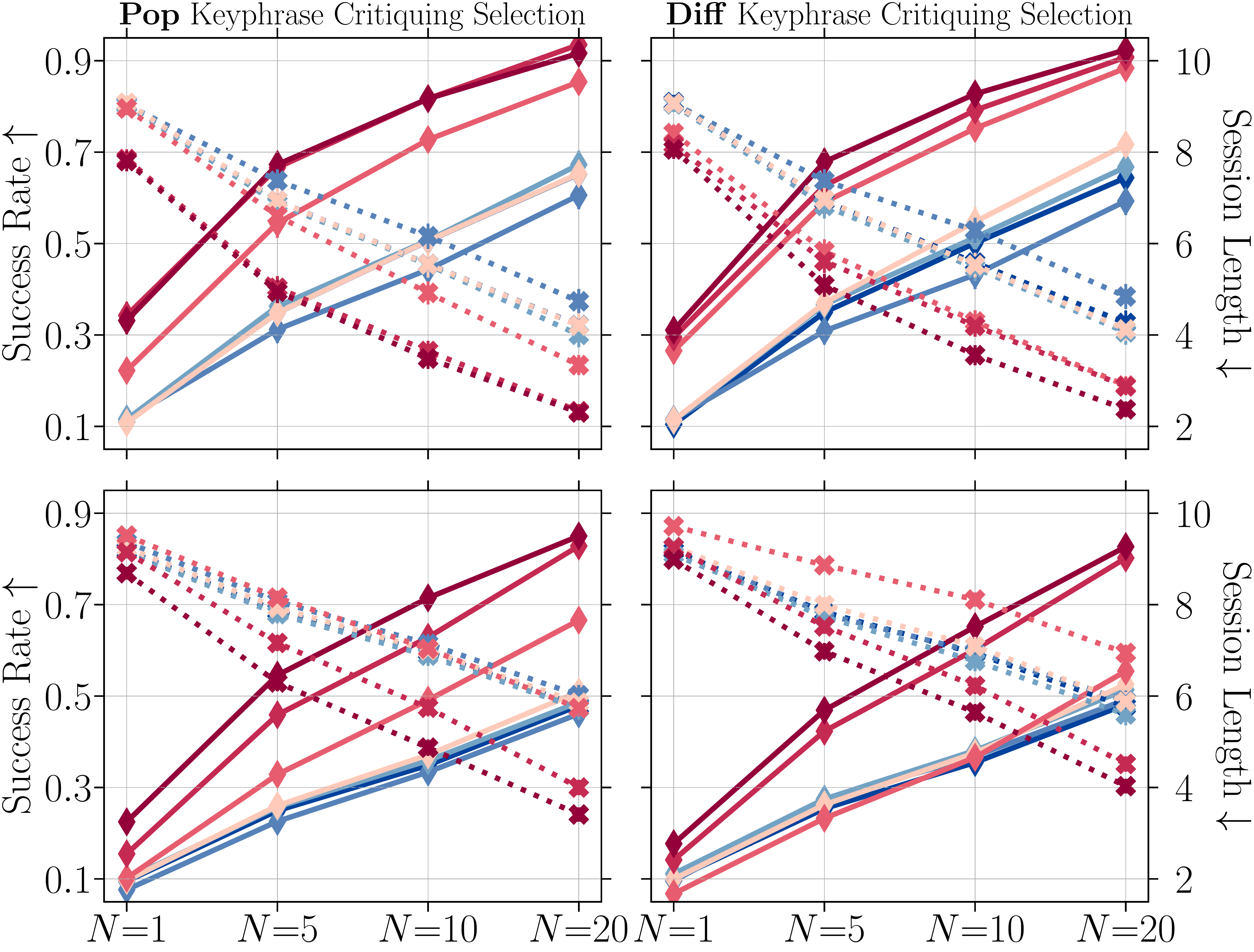}
    \caption{\label{fig_neg_critiquing}Negative critiquing.}
\end{subfigure}
\caption{\label{fig_multi_crit}Multi-step critiquing performance after 10 turns of conversation. We simulate (a) positive and (b) negative critiquing. We report the average success rate (left y-axis) and session length (right y-axis) at different Top-N with 95\% confidence interval.}
\end{figure*}

\subsection{RQ 2: Multi-step Negative and Positive Critiquing Performance Comparison}
\label{sec_rq2}
\subsubsection{Baselines}
We use \textbf{CE-VAE}, which learns an inverse mapping between a critique and the latent space. During critiquing, it averages the user embedding with the critique embedding. We also employ uniform average critiquing (\textbf{UAC})~\cite{luo2020b}, in which the user and critique embeddings are averaged. Finally, we include \textbf{LLC-Score} \cite{luo2020b} and \textbf{LLC-Rank}~\cite{hanze2020}, which extend PLRec~\cite{sedhain2016practical} to co-embed keyphrases into the user embedding. Then, they compute~a weighted average between the user's and each critique's embedding. Weights are optimized via linear programming; LLC-Score uses a margin-based objective, and LLC-Rank uses a ranking-based one. All models were designed for negative critique. For positive critiquing scenarios, we adapt their inner workings accordingly.

\subsubsection{User Simulation} Following prior work~\cite{luo2020b,hanze2020,fast_critiquing}, we run a user simulation to assess the models in a multi-step conversational recommendation scenario. We limit the critiquing iterations to a maximum of 10 turns. The conversations stop if the target item appears within the top-N recommendations on that iteration.
The simulation includes all users and follows Alg.~\ref{alg_dataset}, with the following differences: \begin{enumerate*}
 \item we consider only target items from the \textbf{test} set, and
 \item on Line 7, $\bhku$ is replaced by $\bkpu$ to guarantee the same sequence of critiques and thus a fair comparison. We assess positive and negative critiquing performance in two separate simulations.
 \end{enumerate*}
 
 For the critique selection, we assume the user selects a keyphrase to critique according to two strategies: \begin{enumerate*}
	\item \textbf{Pop}: the most popular keyphrase and
	\item \textbf{Diff}: the keyphrase that deviates the most from the target item keywords. We compare the top recommended items' keyphrase frequency to the target item's keyphrases and select the keyphrase with the largest frequency differential.
\end{enumerate*}

\subsubsection{Multi-Step Critiquing Performance}  
We assess the models with the average success rate and session length at different Top-N. For each user and target item, we sample 299 unseen items, as \cite{fast_critiquing}.

Fig.~\ref{fig_pos_critiquing} shows the results for positive critiquing. Impressively, M\&Ms-VAE\textsuperscript{$+$} significantly outperforms all methods on both metrics and datasets. It confirms that positive critiques are effectively embedded. On the Yelp dataset, M\&Ms-VAE\textsuperscript{3} and M\&Ms-VAE obtain success rates similar to those of the baselines but the worst~session lengths. However, UAC and LLCs clearly underperform~on the Hotel dataset. These results validate our intuition that representing negative and positive critiques via the same encoder is suboptimal. 

Regarding the depiction of negative critiquing in Fig.~\ref{fig_neg_critiquing}, there~is a large margin between M\&Ms-VAE-based models and baselines. 
 Remarkably, the new positive critiquing terms in the loss~function of Eq.~\ref{eq_ssc_pos_neg} dramatically improve the success rate of M\&Ms-VAE\textsuperscript{$+$} and M\&Ms-VAE\textsuperscript{3} compared to M\&Ms-VAE. Also, M\&Ms-VAE\textsuperscript{$+$}~obtains 

\noindent session lengths significantly 
shorter than those of M\&Ms-VAE\textsuperscript{3}.

These observations highlight the importance of inferring the representations of positive and negative critiques differently. Adding users' keyphrase-usage dislikes $\bknu$ as a third modality hinders the positive critiquing performance compared to M\&Ms-VAE due to the generative model $p_{\Theta_{k^-}}$ ,that perturbs its training. Finally, M\&Ms-VAE\textsuperscript{$+$} remedies this problem and produces the best results overall.

\section{Conclusion}

We present M\&Ms-VAE\textsuperscript{$+$}, an extension of M\&Ms-VAE that enables positive and negative multi-step critiquing. The novelty relies on modeling users' keyphrase-usage dislikes as~a third modality while dropping its generative model, an approach made possible by the multimodal factorization. This enables the model to learn powerful positive and negative critique representation. The~key results~show that M\&Ms-VAE\textsuperscript{$+$} matches or exceeds M\&Ms-VAE in recommendation and explanation and is the first model to obtain substantially better positive and negative multi-step critiquing performance.

\bibliographystyle{ACM-Reference-Format}


\end{document}